\documentclass[pre,superscriptaddress,twocolumn,showpacs,preprintnumbers,amsmath,amssymb]{revtex4}
\input psfig.sty
\begin{document}
\title{Equilibration times in numerical simulation of structural 
glasses: Comparing parallel tempering and conventional molecular dynamics}

\author{Cristiano De Michele}
\affiliation{Dipartimento di Scienze Fisiche and INFM, 
 Universit\'a di Napoli "Federico II",
 Via Cinthia (Monte S. Angelo) Ed. G, I-80126 Napoli, Italy}
\author{Francesco Sciortino}
\affiliation{Dipartimento di Fisica, INFM and 
INFM Center for Statistical Mechanics and Complexity,
 Universit\'a di Roma "La Sapienza", 
Piazzale Aldo Moro 2, I-00185  Roma, Italy}

\date{\today}
\begin{abstract}
Generation of equilibrium configurations is  the major obstacle 
for numerical investigation of the slow dynamics in supercooled 
liquid states.
The parallel tempering (PT)
technique, originally proposed for the numerical equilibration of
discrete spin-glass model configurations, has recently been
applied in the study of supercooled structural glasses.
We present an investigation of the ability of
parallel tempering to properly sample the liquid
configuration space at different temperatures, by 
mapping the PT dynamics into the dynamics of the
closest local potential energy minima (inherent structures).
Comparing the PT equilibration process with the 
standard molecular dynamics equilibration process we find that
the PT does not increase the speed of equilibration 
of the (slow) configurational degrees of freedom.
\end{abstract}

\maketitle
%
\label{sec:intro}

As a liquid is cooled below its melting temperature $T_m$
(supercooled liquid) the 
structural time $\tau$
increases considerably. In a small temperature interval, $\tau$ 
changes by more than 13 order of magnitude.
When $\tau$ reaches values bigger than $100 s$ 
the liquid behaves as an amorphous solid, i.e. a  glass. 

In recent years, a considerable interest has been devoted to
the study of the supercooled state of matter, both
theoretically\cite{goetze,parisi,pablo},
experimentally \cite{Angell,goetzepisa,cummins} 
and numerically \cite{lepo,kob,sciortino,harrowell}.
Both thermodynamics \cite{parisi} and dynamic \cite{goetze} 
theories have been proposed  
to explain the rich phenomenology of glassy systems.
Molecular dynamics (MD) simulations have proved to be a powerful
tool for studying simple models for liquids in supercooled states
(for a review, see \cite{kobrev}). Simulation stretching in the
$ns$ time window have
offered the possibility of a detailed comparison between 
theoretical predictions and ``exact'' numerical results.
So far, such comparisons have been limited  to 
weakly supercooled states, i.e. to the temperature region 
where characteristic times are at most of the order of 10
ns. In this region, mode coupling theory (MCT) has shown its
ability in correctly predicting the numerical 
results \cite{kobgleim,sciortino,letz} even for network forming 
liquids \cite{water,sio2prlgen2001}.

The analysis of numerical data has been also very fruitful in
the study of the potential energy surface (PES) --- the so--called 
energy landscape --- of several models\cite{pablo}.
These studies have provided evidence that in equilibrium
 the average basin depth $e_{IS}(T)$ is a 
decreasing function of $T$ \cite{sastrynature}. The number of 
explored local PES minima, commonly named inherent structures (IS) 
 --- the exponential of the configurational entropy
in the inherent structure formalism \cite{stillinger,
sciortinoentropyprl,sastrynature2} --- 
decreases also on cooling. Numerical studies
on aging liquids \cite{epl,jpcm2001} have shown
that the equilibration process is related to the slow search for
deeper and deeper basins on the potential energy surface.  
In the PES framework at least two
different factors control the equilibration time scale: 
(i) the timescale
for escape from a selected basin (a timescale depending on
the kinetic energy) and (ii) the timescale for finding deeper basins
(a time scale depending on the number of accessible basins). 
Which of the two factors is the leading one is 
still an open question. 

Presently, the interesting region where dynamics slow down beyond the
$ns$ time scale can not be studied numerically since the generation of
equilibrated configurations requires prohibitive computational
times. The possibility of disposing of equilibrium configuration 
could
open the possibility of studying, if not the entire structural
relaxation process, at least the initial part of it, where several
interesting phenomena related to the dynamics in disordered structures
are taking place \cite{parisi2prl2001,goetze2pre,ruoccoprl,bosonpeak}.

Several algorithms have been developed 
to improve the equilibration times in numerical simulations 
of glassy systems 
\cite{pivot-nature,pt}. A  study by
Kob and Yamamoto suggests that the parallel tempering (PT) 
may become an important tool to provide independent
equilibrium configurations for structural glasses.
The PT technique \cite{pt} was
developed for dealing with the slow dynamics
of disordered spin systems.
The PT algorithms simultaneously 
simulates a set of $M$ identical 
non-interacting replicas of the system, each of them at a
different $T$. Pairs of replica swap their temperatures 
according to a Monte Carlo procedure.
The basic idea is that each replica performs a
random walk among the $M$ different $T$.
Hence, when the replica explores the high T states,
the probability to escape from its basin is enhanced.

Each of the $M$ replicas, composed by $N$ atoms,  
is described by an Hamiltonian 
\begin{eqnarray}
  H_m({\vec q}_m,{\vec p}_m) = \sum_{i=1}^{N} \frac{1}{2m} 
{\vec p}_i^2 + \Lambda_m(t) E({\vec q}_m) + \nonumber\\
\frac{1}{2} Q 
\left (\frac{\dot s_m}{s_m}\right )^2 + 
\frac{(3N-3)}{k_B T_0} \ln(s_m)
\label{eq:PTHamil}
\end{eqnarray}
where $ E({\vec q}_m)$ is the potential energy of the system.
$\Lambda_m(t)$ is a scaling parameter for the potential energy,
which effectively sets the temperature $T$ of the $m$-th replica
to the value $T_0/\Lambda_{m}(t)$, where $T_0$ indicates the lowest
studied temperature. Consequently the values 
$\Lambda_m(0)$, for $m=0....M-1$ set the 
$M$ different temperatures of the $M$ 
replicas \cite{pt} at time $0$.
The degree of  freedom $s_m$ (last two terms in Eq.\ref{eq:PTHamil}) 
are relative to the Nos\`e thermal bath \cite{nose}. The thermostat 
constrains the average kinetic energy 
of each replica to the value $3/2N k_B T_0$.

The whole Hamiltonian is then:
\begin{equation}
  H = \sum_{m=1}^{M} H_m.
\end{equation}

As discussed in detail
in Ref.\cite{pt} and \cite{KobYam}, 
the choice of  $\Lambda_m(0)$
must guarantee a
significant overlap in the energy distributions
of different replicas, a requirement which oblige to keep $M$
proportionally to the system size.

In this article we focus on  the time requested to 
find the low inherent structure 
configurations visited in equilibrium.
More specifically, to evaluate if  
the PT technique is a viable candidate to equilibrate structural glasses, we compare the PT 
and the conventional MD dynamics by computing the
inherent structure energy as a function of the simulation time. 
Since $e_{IS}$ is a much more sensitive indicator of
equilibrium than the 
total potential energy, we can 
put the PT technique under stringent test.

\section{MODELS AND DETAILS OF SIMULATION}
\label{sec:model}

The system we investigated is the monoatomic Lennard-Jones (LJ) 
model modified by adding a many-body anti-crystalline potential designed to
inhibit crystallization \cite{angelani}.  The $\epsilon$ and $\sigma$
parameter of the LJ potential are chosen as unit of energy and length
respectively. The LJ potential is truncated and shift at $2.5$.  The
potential energy, which includes the
anti-crystalline potential is

\begin{eqnarray}
 E({\vec q}_m) &&= V_{LJ}({\vec q}_m) + \nonumber\\ &&\frac{1}{2}\alpha
\sum_{\vec k} 
 \theta(S_m(\vec k) - S_0) [S_m(\vec k) - S_0]^2.
\label{eq:EMLJ}
\end{eqnarray}

where $V_{LJ}$ is the LJ part of the potential and the sum is over all 
$\vec k$ such that $k_{max}-\Delta k<\|\vec k\|<k_{max}+\Delta k $. The other 
terms in Eq.[\ref{eq:EMLJ}] set in only when any of the 
wavevector around the structure factor peak increases beyond the
value $S_0$ and act by damping the unwanted 
crystalline like density fluctuation.
We chose a number density $\rho=1$,
and $S_0 = 10$,
$k_{max}=7.12$, $\alpha = 0.83$ and $\Delta k=0.34$ for the anti-crystalline parameters
as proposed in Ref.\cite{angelani}.  
The integration timestep is
$0.0025$, in time units of $\sqrt{m \sigma^2 / \epsilon}$.  
The dynamics for this model has been previously studied \cite{newangelani}. It has been shown that a fast decrease of the structural times takes place below $T=1$. The $T$ dependence of $\tau$ follows
a power law  in $T-T_x$, with $T_x \approx 0.475$. $T_x$ has been identified with the ideal MCT for this model, an hypothesis supported also by an analysis of the $T$ dependence of the
diffusive directions \cite{saddles}.

The PT algorithm is identical to that encoded in Ref.\cite{KobYam} and we
refer to that paper for details on the technique.  
The algorithm we implement uses
$M=14$
identical non-interacting replicas each composed of $N=256$ 
particles. The $14$ temperatures are chosen to span an
range from $T=1.05$ down to $0.485$, in particular the temperatures we used
are the following: $0.485$, $0.518$, $0.534$, $0.562$, $0.597$, $0.646$, 
$0.694$, \allowbreak $0.745$, $0.80$, $0.85$, $0.90$, $0.95$, $1.0$, $1.05$.

The Hamiltonian of one replica is that of Eq. [\ref{eq:PTHamil}].
All replicas evolve according to the standard Nose' constant temperature MD
simulation.  Every $1000$ steps an attempt to exchange the scaling
parameter of all pair of replicas with adjacent temperatures
(swap of the $\Lambda$ values) is
performed using the following criterion: an exchange is accepted in
Metropolis fashion, i.e. the acceptance ratio is:
\begin{displaymath}
w_{m,n} = \left\{
\begin{array}{ll}
1\ , & \Delta_{m,n}\leq 0 \\
\exp(-\Delta_{m,n})\ , & \Delta_{m,n} > 0
\end{array}\right.
\end{displaymath}
where $\Delta_{m,n} = \beta_0 (\Lambda_n - \Lambda_m) [E(q_m) - E(q_n)]$.
The events with $i=0,2,4,\dots$ or $i=1,3,5,\dots$
are repeated alternatively every $1000$ integration steps.  

The outcome of such calculation are, in principle, equilibrium 
configurations in the canonical ensemble at the $M$
different temperatures.

To estimate the time requested to the PT algorithm to 
equilibrated all
replicas we start our PT algorithm with $M$ replicas
 extracted from a
previously generated ensemble of equilibrium configurations at
$T=1.05$. At this $T$, the structural relaxation time is of the order
of 1000 steps and hence generation of equilibrium configurations with
conventional MD does not pose any problem. By starting with this
ensemble of configurations, the PT equilibration time is by
construction $0$ for the highest temperature.  
We performed $20$ of such
independent PT runs to improve the statistic. Each of such run
lasts two million time step. Hence, in the PT part of the work, the
equation of motion have been integrated more than
40 million times.

The same starting configurations ($T=1.05$) are also used as initial
configurations for conventional constant temperature MD simulations at
several bath temperatures $T$, to compare the rate of equilibration of
PT and conventional MD algorithms.
For each temperature of these MD simulations we performed $16$ 
independent runs to improve statistics. 

Local minima configurations have been calculated
via conjugate gradient minimization. The minimization process is
considered completed when the potential energy change associated with
one iteration is less than $10^{-15}$ to ensure a great accuracy.

\section{RESULTS}
\label{sec:results}

We focus here on the evolution in time of the
inherent structure energy $e_{IS}$, comparing the PT and MD 
procedures.  Recent work \cite{epl,mossacondmat,jpcm2001} has provided
evidence that following a $T$ change, the system response is 
characterized by two different time scales. A short time scale, related to the
equilibration of the system within a well defined basin of the
energy landscape and a slow time scale related to the search for
basin of the ``right'' depth. The evolution of one-time quantities 
carries on information on these two time scales. 
After a sudden change of $T$, 
a very fast decreases of the vibrational energy --- corresponding to the fast equilibration to the new bath $T$ of the 
intra-basin vibrational motions --- is observed. This 
fast change is 
followed by a much slower decrease which corresponds to the 
slow decrease of the basin's depth.
The absolute change in $e_{IS}$ during the aging process is significantly
smaller than the change in the total potential energy 
and hence it requires a careful analysis to be detected.
A way which has been proven fruitful to separate 
large fast component and small slow component is 
to monitor directly the evolution of $e_{IS}$. Building on the 
expertise developed in recent years, we adopt this indicator
as effective tool to monitor the equilibration of the system
in configuration space. 

We note on passing that, since $e_{IS}$ is a small component 
of the potential energy --- being the intra-basin 
vibrational part dominating --- a nice scaling of the 
total potential
energy distribution  may not guarantee perfect equilibration.

\begin{figure}
\begin{center}
\psfig{file=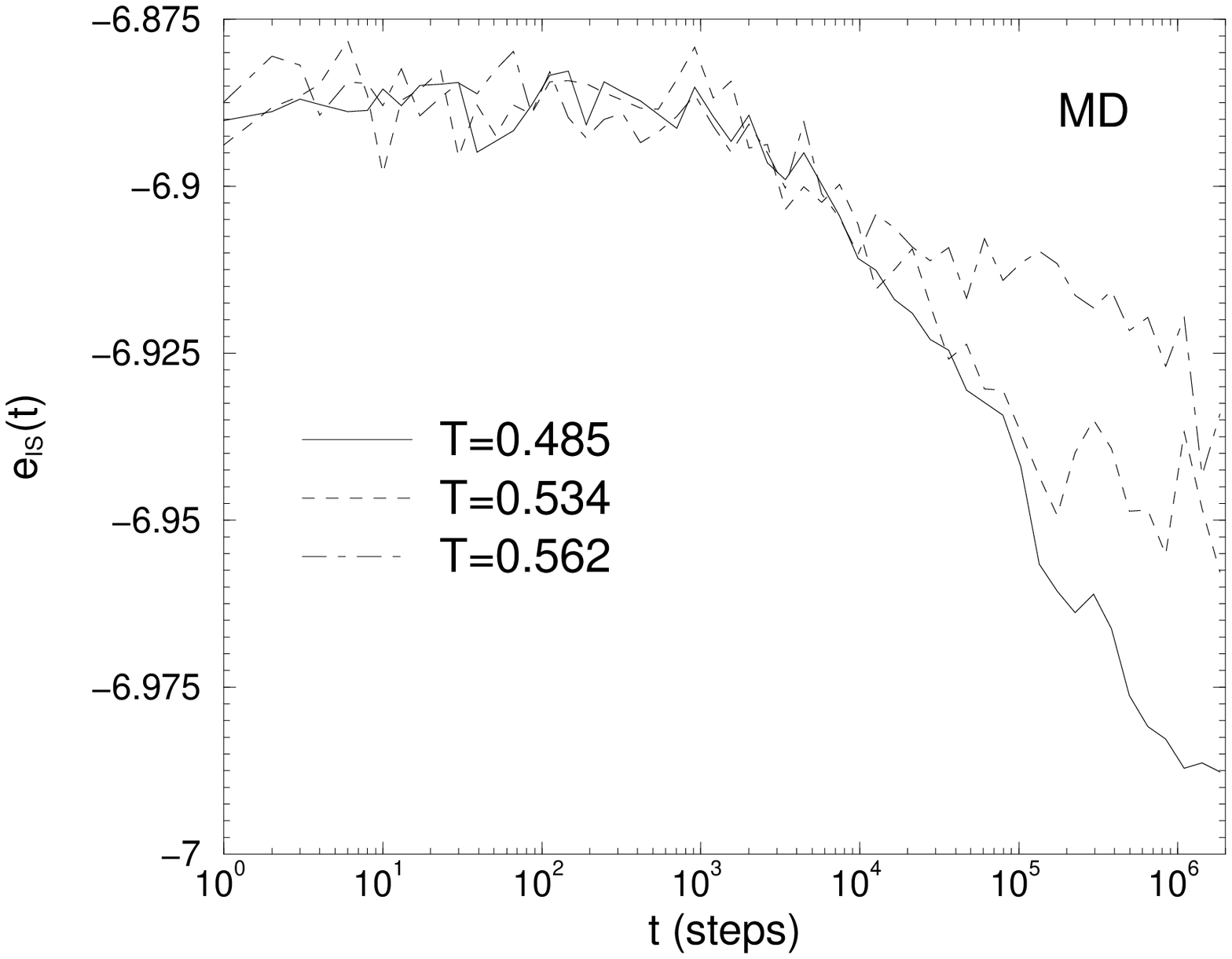,width=8cm,height=6cm}
\psfig{file=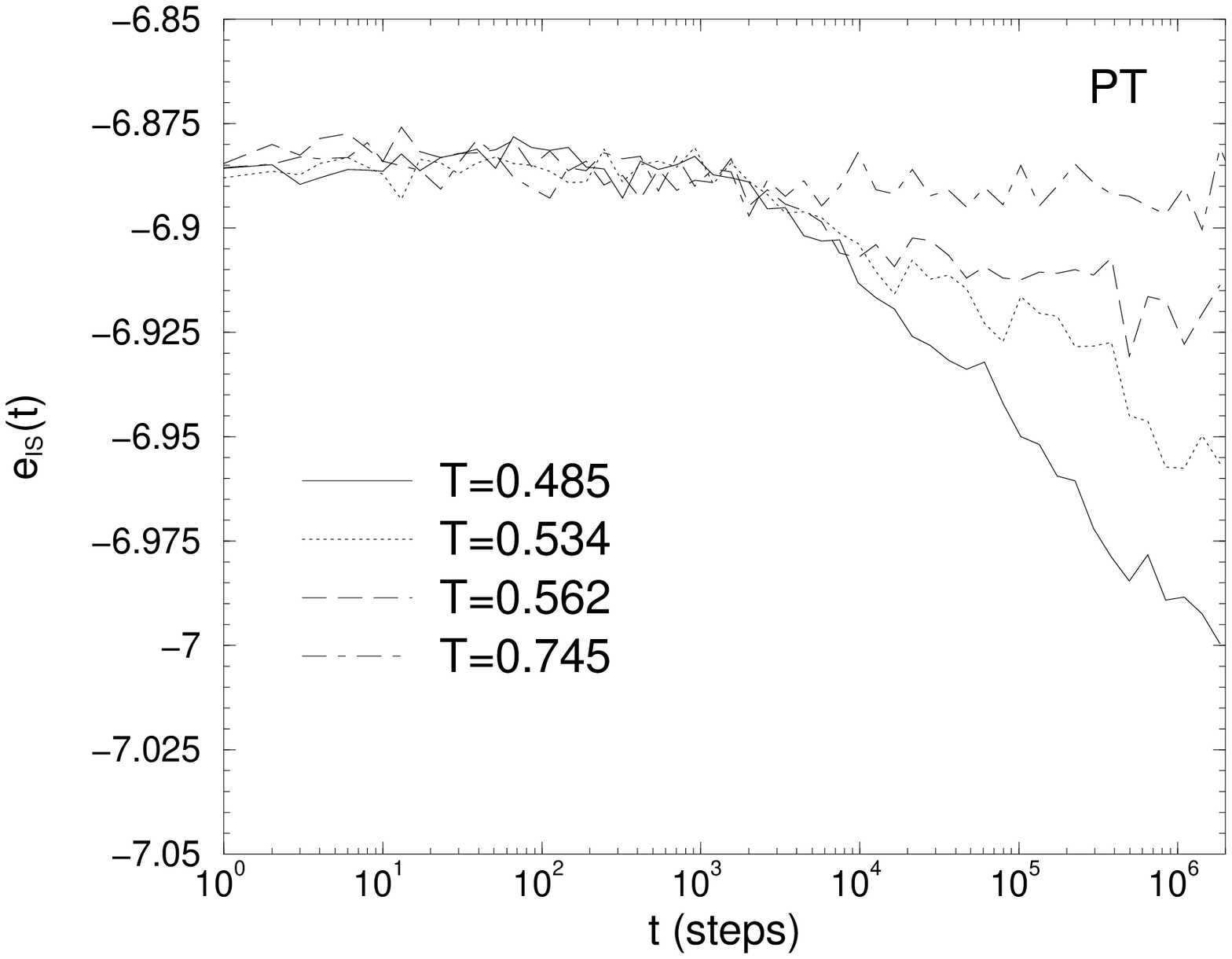,width=8cm,height=6cm}
\end{center}
\caption{Inherent structure energies as a function of time for MD (Top) and PT (Bottom).}
\label{fig:botheis}
\end{figure}

Figure \ref{fig:botheis} compares the time evolution of the
inherent structure energy in the conventional
MD (Top) and in the PT (Bottom) runs. In both cases, by construction, 
the initial $e_{IS}$ coincides with the
equilibrium value of $e_{IS}$ at $T=1.05$. As time goes on, 
each replica starts to explore larger and larger parts of the 
configuration space selecting configuration with lower and lower
values of $e_{IS}$. The equilibration process lasts until the
equilibrium value of $e_{IS}$ is reached. The same picture applies to
the conventional MD case, where the simulation indeed reproduces
the aging  process following a $T$ jump from $T=1.05$ to 
the new bath temperature.

\begin{figure}
\begin{center}
\psfig{file=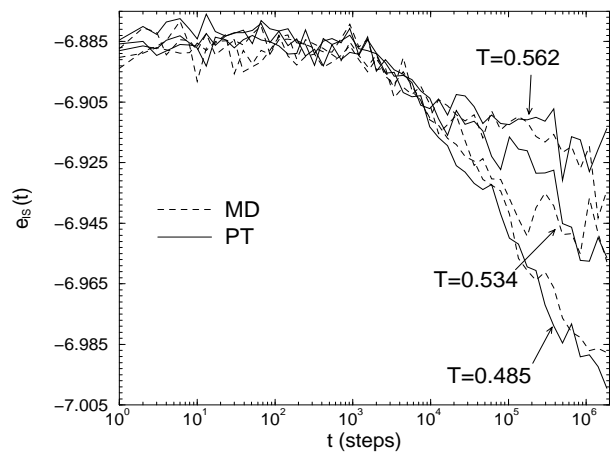,width=8cm,height=6cm}
\end{center}
\caption{Inherent structure energies as a function of time: comparison between MD and PT .}
\label{fig:eiscomp}
\end{figure}

Fig. \ref{fig:eiscomp} compares 
the time evolution of $e_{IS}$ for PT and MD
simulations at three different temperatures.
In all cases, we find clear indications that 
the equilibration of the slow degrees of freedom 
does not depend on the procedure adopted. 

\section{DISCUSSION}
\label{sec:concl}
The data shown in Fig.\ref{fig:eiscomp} very clearly show that
for the Lennard Jones case investigated in this article
no improvement in equilibration rates is achieved by implementing
the PT algorithm. 

For the related model of binary mixture
Lennard Jones \cite{kobrev}, the number of basins with depth $e_{IS}$, 
in the range of $e_{IS}$ values characteristic of the 
PES region explored above $T_x$  is well represented by
a gaussian distribution. 
The total number of basins has been shown to scale
with the size $N$ of the system as $e^{\alpha N}$ with
$\alpha\approx 0.8$ \cite{sciortinoentropyprl}. For our 256 atom case, this correspond to
about $10^{87}$ basins. Under such complicated potential energy landscape conditions, 
an unbiased search for the location of the deepest basin would require
an order of
$10^{87}$ attempts! In this respect, it is possible that
the rate of equilibration at low temperatures is significantly
controlled by a simple entropic effect. This could explain why 
the possibility of overcoming barriers with higher probability
offered by PT does not favor a faster equilibration process.
This picture is also consistent with the fact that in the
$T$-region explored (which is still above $T_x$) saddle
dominated dynamics is dominant. Recent instantaneous normal
 modes analysis  \cite{normalmodes}
has indeed provided evidence that above $T_x$ the 
system explore mostly regions of the potential energy landscape 
which are characterized by a large number of negative curvature
directions. No activated processes are 
requested in this condition to change local basin.

It would be interesting to find out if the PT algorithm may be
valuable in studying strong liquids, for which less relevant 
changes in
the PES are taking place on cooling as compared to fragile
liquids \cite{saikevoivod-nature} and for which activated 
processes are
dominant at low $T$.  Preliminary indications \cite{kobprivatecom}
seems to suggest that this may be the case.
It would also be important to correlate the efficiency of
PT with the structure of configuration space and connectivity 
between distinct potential energy surface basins. A possible 
line of research 
could be to compare  $e_{IS}(t)$ for PT and MD in clusters 
with different disconnectivity graph \cite{disconnectivity}
types.

The fact that in the explored $T$-region equilibration times
are not improved in the PT case is consistent with recent finding
in the field of protein simulations.
This analogy is not at all surprising due to the significant
similarities in the problem of glass formation and 
protein folding \cite{walesscience}. Again, the similarity points 
to the structure
of the PES as key element in the control of the 
characteristic times.

To conclude,
we like to call the reader attention on the fact that
to study a fixed $T$-range, the PT technique requires a
number of replicas which increases linearly 
with the system size, to 
guarantee proper overlap in the potential
energy distributions of adjacent replicas 
and hence a significant replica exchange rate. Moreover, all
replicas have to be simulated for the same total time interval.
This time is fixed by the lowest temperature, which is 
characterized by a relaxation time which may well be several order
of magnitude smaller than the one of the top temperature.
Both effects concur in making PT a not very convenient algorithm
for simulating structural glasses as compared to conventional MD.
Indeed, in MD, the total simulation time at each temperature
can be chosen to scale with the structural relaxation time.
Of course, bookkeeping facilities are enhanced in the PT case.

\acknowledgments
The authors warmly thank Walter Kob for useful discussions.
We acknowledge support from INFM - Iniziativa Calcolo Parallelo
and INFM PRA HOP.

\end{document}